\documentclass{appolb}
\usepackage{graphicx}
\usepackage{hyperref}
\usepackage{url}

\begin{document}
\eqsec  
\title{Modeling NNLO jet corrections with neural networks%
\thanks{Cracow Epiphany Conference 2017 proceedings. Preprint: CERN-TH-2017-076}%
}
\author{Stefano Carrazza
\address{Theoretical Physics Department, CERN, Geneva, Switzerland}
}
\maketitle
\begin{abstract}
We present a preliminary strategy for modeling multidimensional
distributions through neural networks. We study the efficiency of the
proposed strategy by considering as input data the two-dimensional
next-to-next leading order (NNLO) jet $k$-factors distribution for the
ATLAS 7 TeV 2011 data. We then validate the neural network model in
terms of interpolation and prediction quality by comparing its results
to alternative models.
\end{abstract}
\PACS{07.05.Mh, 12.38.-t}
  
\section{Introduction}

The calculation of higher order corrections to QCD processes, usually
related to LHC measurements, requires intensive computing power and
time. However, higher order corrections, such as the next-to-next
leading order (NNLO), are essential in many circumstances,
\textit{e.g.}~during the determination of parton distribution
functions (PDFs)~\cite{Ball:2014uwa,Ball:2012cx} where theoretical
predictions are computed several times during the minimization
strategy adopted by the PDF fitter.

Nowadays this performance limitation is overcome by the inclusion of
NNLO corrections in PDF fits through the pre-evaluation of
$k$-factors, defined as the ratio between predictions at NNLO and NLO
computed with the same set of PDFs. In a NNLO PDF fit the $k$-factors
are multiplied to the respective NLO predictions which are evaluated
using high performance techniques, such as {\tt
  APPLgrid}~\cite{Carli:2010rw} or {\tt
  APFELgrid}~\cite{Bertone:2016lga}, obtaining acceptable convolution
timings for minimization algorithms.

The aim of this proceedings is to present a preliminary strategy based
on neural networks and backpropagation to build a model for the
$k$-factors multidimensional distribution, \textit{i.e.}~$k$-factors
obtained from the ratio of differential distributions. There are two
basic motivations for modeling $k$-factors. First, the possibility to
provide a reliable method to interpolate and extrapolate $k$-factors
for a specific process in a custom kinematic range. This is
particularly useful when the computation of the exact $k$-factor
requires several computing hours. Secondly, if the procedure is
unbiased we have the possibility to estimate the reliability of the
input $k$-factors, in terms of central values and uncertainties, by
looking at the fit quality. In simple words, if an unbiased fit does
not converge and produce poor statistical estimators there is a high
probability that data and its uncertainties are inconsistent.

In the next sections we start by presenting the input data and
strategy used here. Then we discuss the model results and validation
estimators. We conclude by showing the behavior of the neural network
model in an extrapolation region and comparing its results to
alternative models.

\section{Building the neural network model}

\subsection{Data}

The target data selected for our exercise is the $k$-factors for the
fully differential jet production at NNLO for the ATLAS 7 TeV 2011
data~\cite{Aad:2014vwa}. These results have been recently published in
Ref.~\cite{Currie:2016bfm} after preliminary studies perfomed in
Refs.~\cite{Currie:2013dwa,Carrazza:2014hra} and consist in a
two-dimensional $k$-factors distribution binned in $(p_T,y)$, the
leading jet transverse momentum and its rapidity. Moreover, this data
is reconstructed with the anti-$k_T$ algorithm with $R=0.4$ and the
kinematic coverage is $p_T=[100,2000]$ GeV with $|y|<3$.

\subsection{Strategy}

Let us consider the full ATLAS 7 TeV 2011 dataset with $n_{\rm
  dat}=140$ points. In terms of notation, for each point
$i=1,...,n_{\rm dat}$ we represent the corresponding $k$-factor as the
pair $\{k_i,\delta k_i\}$ where $k_i$ is the $k$-factor central value
and $\delta k_i$ its uncertainty. The aim of the strategy proposed
here is to determine a neural network model $y_{\rm
  pred}=\mathcal{N}(p_T,y)$ which minimizes the loss function:
\begin{equation}
  \chi^2 = \sum_{i=1}^{n_{\rm dat}} (k_i-y_{{\rm pred},i}) \sigma_{ij}^{-1}
  (k_j-y_{{\rm pred},j}),
\end{equation}
where $\sigma_{ij}$ is the covariance matrix constructed from the
$\delta k_i$ terms, which in our case is a simple diagonal matrix due
to the lack of information on extra sources of correlations.
The fitting procedure is then summarized by the following steps:
\begin{itemize}
\item We generate artificial Monte Carlo replicas from the original
  $k$-factor input data, by following the procedure adopted in the
  NNPDF framework~\cite{Ball:2014uwa}. This bootstrap procedure
  produces data replicas where central values are shifted following
  random Gaussian noise proportional to the uncertainties stored in
  the covariance matrix. This mechanism provides a simple way to
  propagate the input data uncertainty to the model, so our final
  model will provide predictions for uncertainties.
\item The data of each MC replica is then randomly divided into two
  groups: training and validation. The training data is used to train
  the neural network through the minimization algorithm, meanwhile the
  validation data is used to control the quality of the fit, avoiding
  overlearning and underlearning.
\item We adopt the stochastic gradient descent controlled by
  backpropagation for the minimization strategy. The stopping
  condition is implemented through the look-back algorithm which
  stores the weights and biases of the neural network at the minimum
  of the validation error function.
\end{itemize}

\subsection{Algorithm settings}
\label{sec:setup} 

We have implemented the above strategy using {\tt Keras
  v1.1.1}~\cite{chollet2015} with {\tt Tensorflow
  v1.0.0}~\cite{tensorflow2015-whitepaper} backend. This choice is
motivated by the great advantages provided by these codes like fast
prototyping and high flexibility when testing several optimization
algorithms and neural network architectures.

The final settings adopted in this exercise have been obtained through
an intensive hyperparameter grid search, where we tested the quality
of the loss function of the training and validation sets over several
configurations. Our current best setup consists in a multilayer
perceptron network with architecture 2-5-3-1 where the hidden layers
have hyperbolic tangent activation functions and the last layer is
linear. We have two input nodes that take pairs of points $(p_T,y)$
and one single output node which represents the $k$-factor
prediction. In terms of training and validation split we divide the
original data into 50\%-50\% random sets for each MC replica. We use
as optimizer the RMS propagation with learning rate of $l_r=0.01$
associated to the look-back stopping algorithm on 100k epochs. Using
this setup a single replica fit usually takes 5 minutes to complete
the minimization in a single core.

Finally, we remove replica outliers by applying a $\chi^2$ veto
condition in which a neural network replica is accepted if its
$\chi^2$ to the original dataset is within 4-$\sigma$ of the average
over all replicas. The results presented in the next sections are
based on a final set of $n_{\rm rep}=500$ replicas.

\subsection{Results and validation}

\begin{figure}
  \includegraphics[scale=0.4]{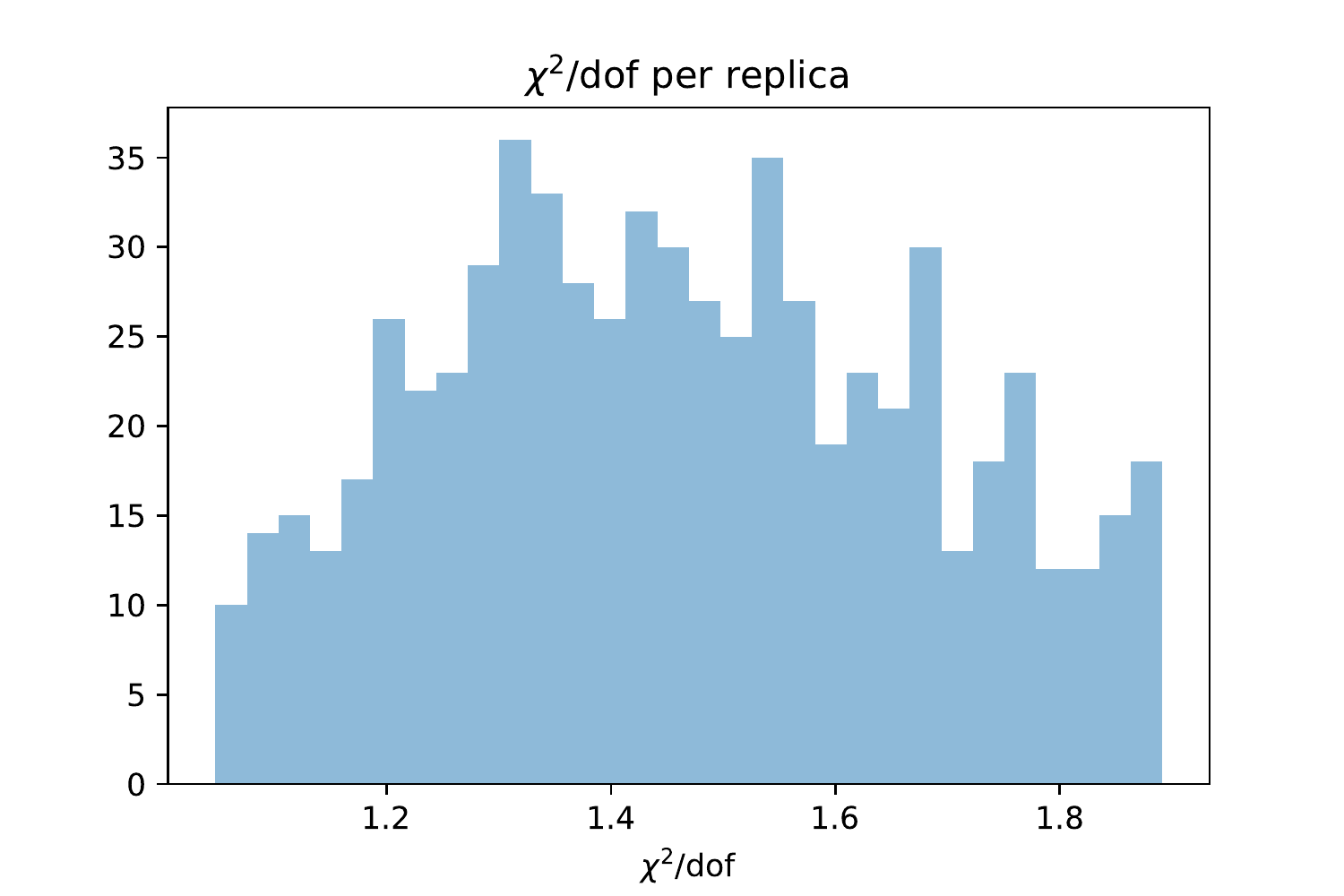}\includegraphics[scale=0.4]{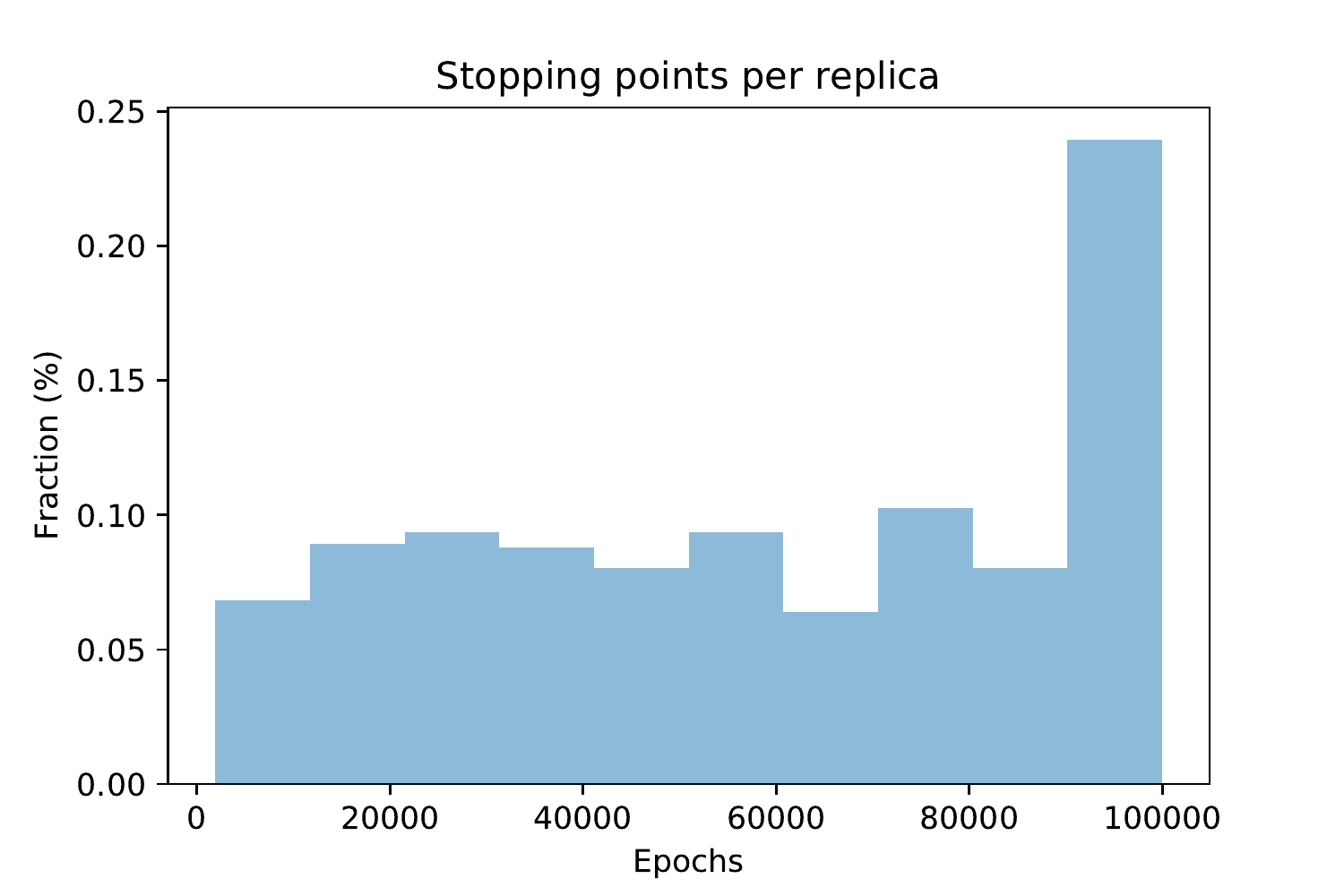}
  \caption{On the left, the $\chi^2$/dof for the full dataset for each
    replica. On the right, the distribution of stopping epoch for each
    replica.}
  \label{fig:figure1}
\end{figure}

In the left plot of Figure~\ref{fig:figure1} we show the distribution
of the total $\chi^2$/dof to the original input data for each neural
network replica. The shape and central value of this distribution
confirm the good quality of the trained model. On the right plot of the
same figure, the distribution of stopping epochs is presented for each
replica. In the current setup, less than 25\% of the total number of
replicas stops at the maximum number of iterations, while the other
replicas stop uniformly between 1k and 90k iterations.

\begin{figure}
  \includegraphics[scale=0.4]{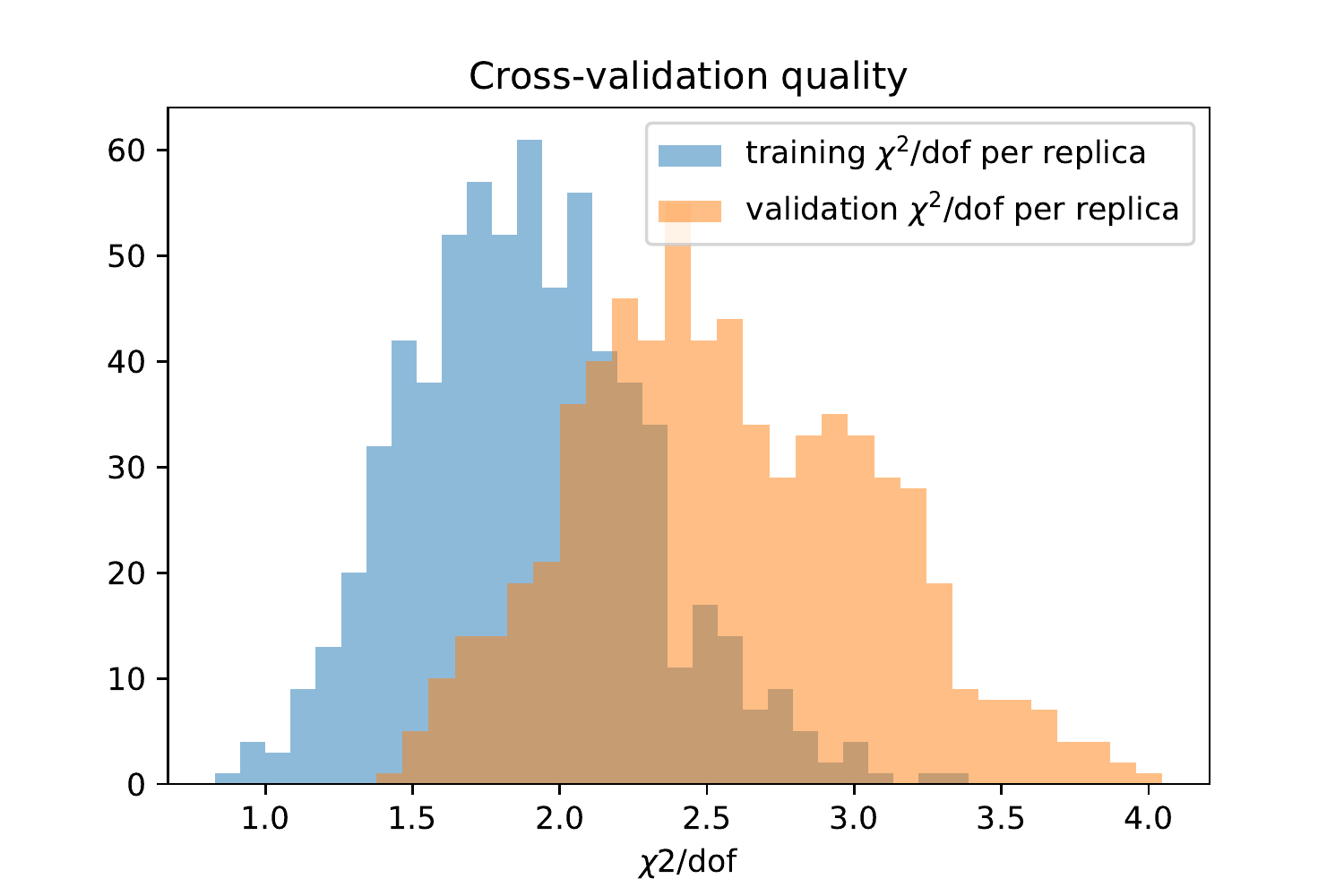}\includegraphics[scale=0.4]{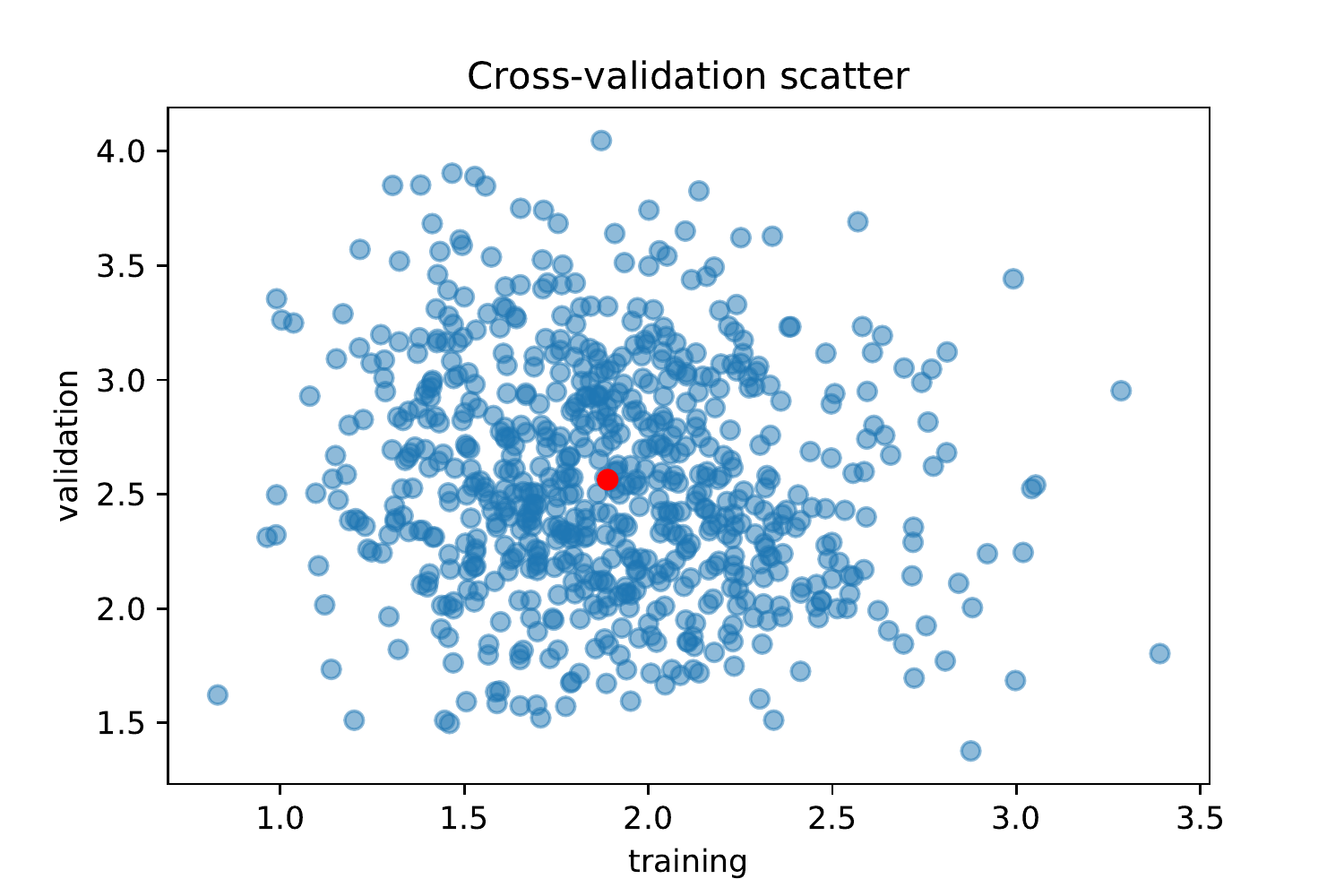}
  \caption{The training and validation $\chi^2$/dof distribution
    histogram (on the left) and scatter plot (on the right). The
    average value for training and validation is shown by the red
    marker in the scatter plot.}
  \label{fig:figure2}
\end{figure}

In Figure~\ref{fig:figure2} we show the training and validation
$\chi^2$/dof distributions for all replicas. The left plot shows the
histogram distribution meanwhile on the right we have the scatter plot
of the same quantities, \textit{i.e.}~the blue points, and the red
marker represents the average value for both quantities. Also here,
the shape and peak values of both distributions are close to each
other and in average within the $\chi^2$/dof interval $[1.8,2.5]$,
which suggests that the model is trained satisfactorily.

\begin{figure} 
  \includegraphics[scale=0.35]{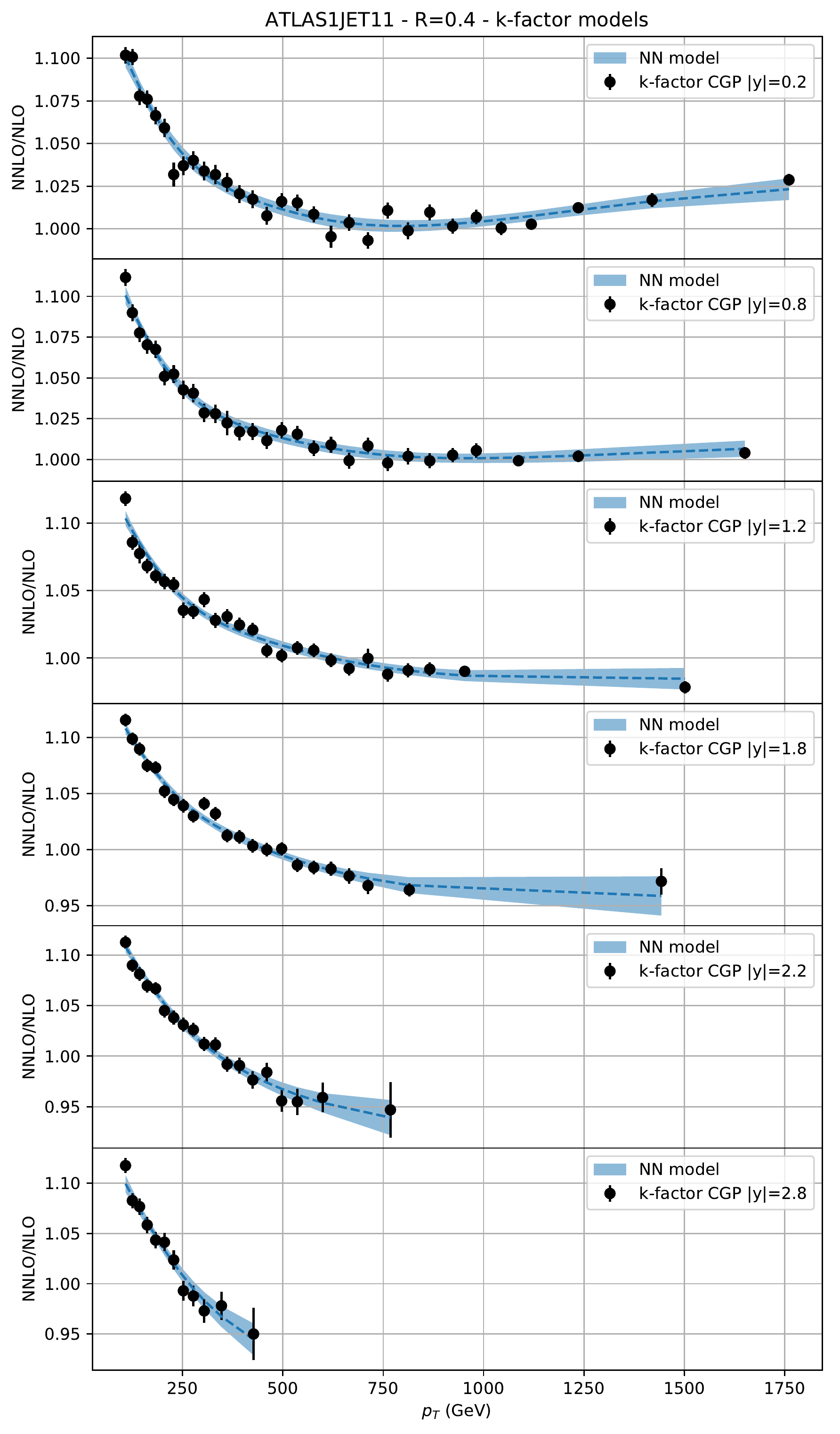}\includegraphics[scale=0.35]{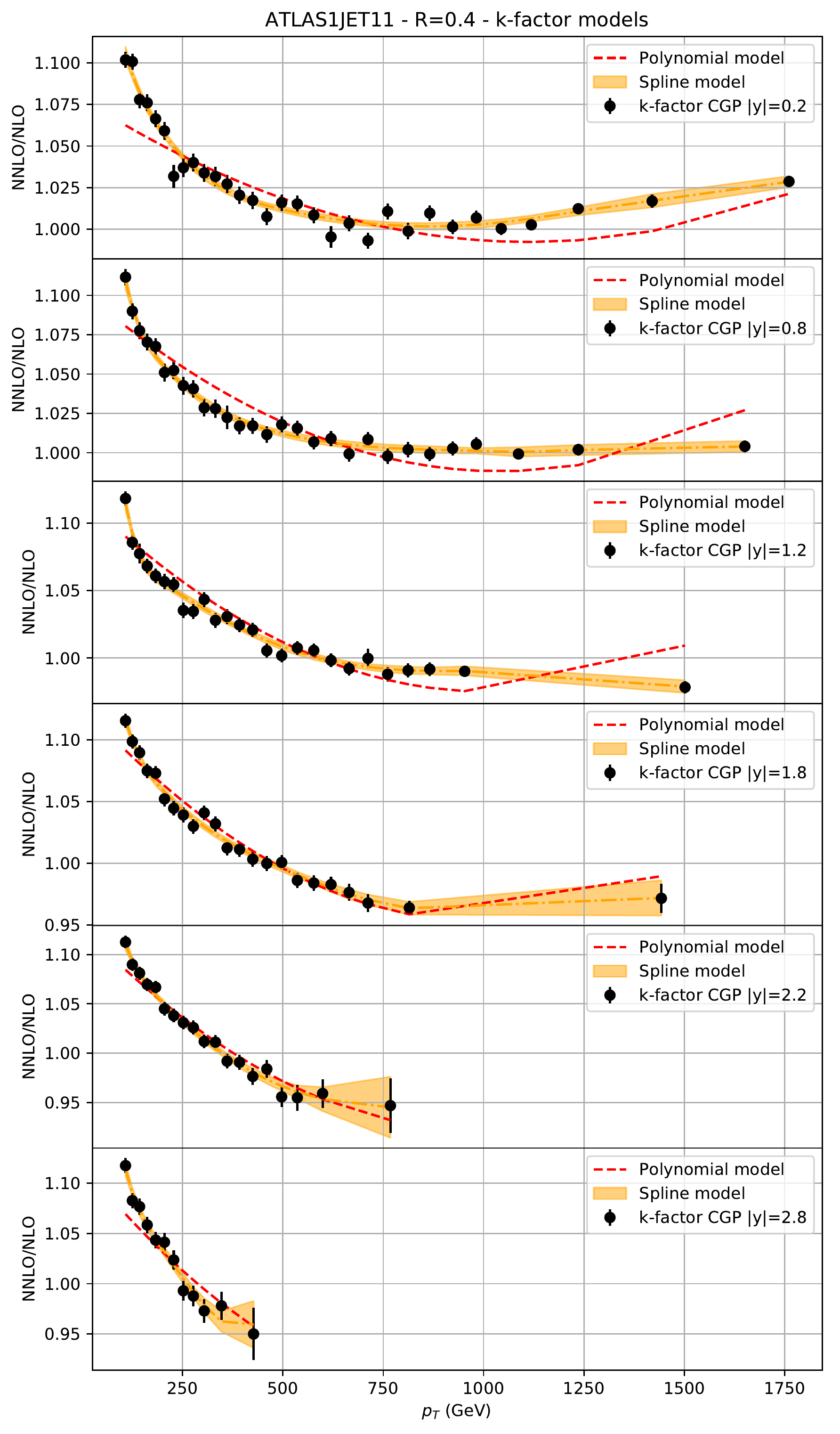}   
  \caption{Model results compared to the input $k$-factors for the
    ATLAS 7 TeV 2011 data. On the left, the $k$-factors (black points)
    separated in bins of rapidity with the neural network model
    central value and uncertainty predictions. On the right, similar
    results for the polynomial (in red dashed lines) and the spline
    (yellow band and line) models.}
  \label{fig:results}
\end{figure}

The left plot in Figure~\ref{fig:results} shows the final results for
the neural network model. The black points are the input $k$-factors,
following the ATLAS 7 TeV 2011 data kinematics, divided by rapidity
bins. The blue line and band are respectively the neural network
central value and uncertainty prediction. From the plots we conclude
that the neural network model is performing well in representing the
shape variation of $k$-factors in the $(p_T,y)$ bins. For comparison
reasons, on the right side of the same figure, we plot predictions for
the same input data using a 2nd degree two-dimensional polynomial
parametrization (red line) and a bivariate cubic spline (yellow band)
both trained with generalized least-square minimization. In
Table~\ref{table:chi2} we computed the total $\chi^2$/dof to the
original input data for each model.

First of all, we conclude that the polynomial model, even if it is
simpler and faster to fit, usually requires a long period of fine tune
of its degree and configuration. We observe that with our input data,
polynomial predictions are not sufficiently compatible with data. On
the other hand, the spline model provides a good interpolation of the
input data, however as we will see in the next section we have to
select a model which is stable over kinematics variations, including
the extrapolation region. Furthermore, it is important to highlight
that the polynomial and spline models will require special fine tunes
when generalizing the problem to higher dimensions, meanwhile for the
neural network model we will have just to increase the number of input
nodes and perform the hyperparameter grid search.

\begin{table}
  \centering{
    \begin{tabular}{l|c}
      \hline 
      \textbf{Model} & $\chi^{2}$/dof\tabularnewline
      \hline 
      \hline 
      Neural-network & 0.93\tabularnewline
      \hline 
      Spline & 0.66 \tabularnewline
      \hline 
      Polynomial & 5.92\tabularnewline
      \hline 
  \end{tabular}}
  \caption{Total $\chi^{2}$/dof for the neural network model
    \ref{sec:setup}, the bivariate cubic spline and the 2nd degree
    two-dimensional polynomial.  \label{table:chi2}}
\end{table}

\section{Extrapolating predictions}

In order to test the extrapolation of all models presented in the
previous section we have computed predictions using the kinematic
settings of ATLAS 7 TeV 2010~\cite{Aad:2010ad}. The only difference
between this dataset and the ATLAS 7 TeV 2011 used to train the models
consists in a wider range and binning of $(p_T,y)$, in particular
$|y|<4.4$ and $p_T=[20,1500]$ GeV. The ATLAS 7 TeV 2011 and 2010 have
exactly the same theoretical setup: $\sqrt{s} = 7$ TeV and $R=0.4$
anti-$k_t$ jet reconstruction radius and algorithm.

\begin{figure}
  \includegraphics[scale=0.35]{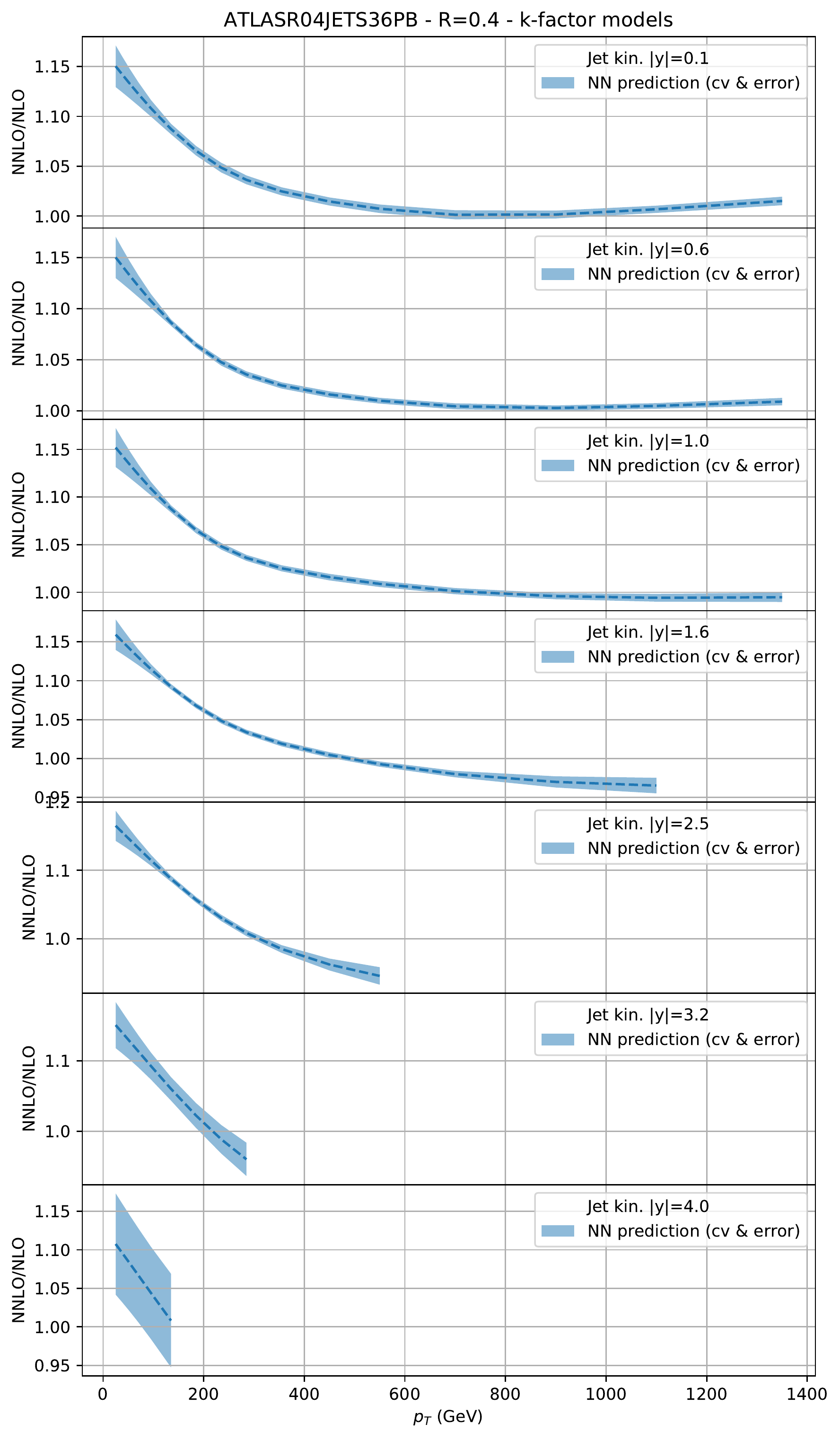}\includegraphics[scale=0.35]{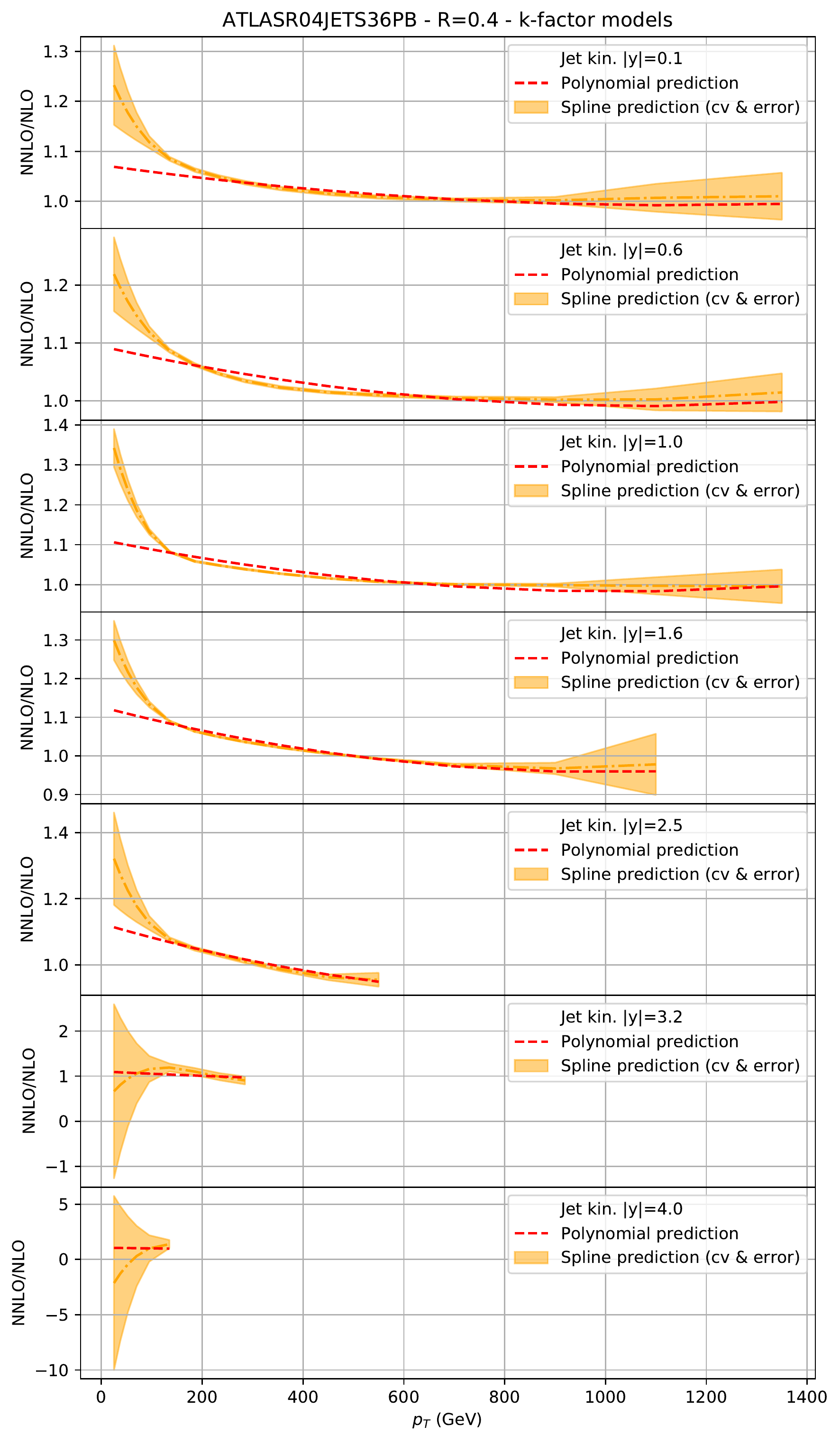}   
  \caption{Same as Figure~\ref{fig:results} for ATLAS 7 TeV
    2010~\cite{Aad:2010ad}.}
  \label{fig:results2}
\end{figure}

Figure~\ref{fig:results2} illustrates the predictions for the neural
network model (left side) and the spline and polynomial (right side)
for the ATLAS 7 TeV 2010 in $(p_T,y)$ bins. As we do not have yet
available the final exact $k$-factors for this experiment, in order to
compute the final total $\chi^2$/dof for each model, we can judge the
quality of extrapolation prediction by looking at central values and
uncertainties provided by each model. The most stable behavior among
all the three different models is the one provided by the neural
network, which simulates the typical shape steepness when increasing
the rapidity and reducing the $p_T$. The spline model fails to predict
such regions, in particular we highlight the presence of negative
predictions and large uncertainties. We also observe that
uncertainties oscillate if the requested $(p_T,y)$ bin is too
different from the original input data.

\section{Conclusions and outlook}

From this exercise we conclude that there are at least two clear
advantages in using neural networks to model multidimensional
distributions: no need for a model definition, neural networks can
simulate every function; the possibility to interpolate and
extrapolate multidimensional distributions from two or higher
dimensions easily.

We outlook that this model could be applied in many distributions
provided by Monte Carlo generators, by at least removing the
requirement of rebinning. We also highlight that the such model could
provide a quantitative measurement of the input data consistency.

S.C. is supported by the HICCUP ERC Consolidator grant (614577).

\end{document}